\documentclass[twocolumn, journal]{IEEEtran} 

\makeatletter 
\def\ps@headings{% 
\def\@oddhead{\mbox{}\scriptsize\rightmark \hfil \thepage}% 
\def\@evenhead{\scriptsize\thepage \hfil \leftmark\mbox{}}% 
\def\@oddfoot{}% 
\def\@evenfoot{}} 
\makeatother 
\usepackage{latexsym,times,epsf,amsmath,amssymb,amsfonts,graphicx}
\usepackage{subfigure}
\usepackage{multirow}
\usepackage{algpseudocode}
\usepackage{algorithm}
\usepackage{mathrsfs}
\usepackage{array}
\usepackage{setspace}
\usepackage{cite}
\usepackage{csquotes}
\usepackage{comment}
\usepackage{subfigure}
\usepackage{cite}
\usepackage{amsthm}
\usepackage{amsmath}

\usepackage{color}
\usepackage{bbm}

\newcommand{\specialcell}[2][c]{%
  \begin{tabular}[#1]{@{}c@{}}#2\end{tabular}}

\renewcommand{\theequation}{\arabic{equation}}
\allowdisplaybreaks
\begin{document}
\title{Study on a Low Complexity ECG Compression Scheme with Multiple Sensors}
\author{
\IEEEauthorblockN{Pengda Huang
%Wenbo Wang,~\IEEEmembership{Student Member,~IEEE} and
%Yiming Pi,~\IEEEmembership{Senior Member,~IEEE}\vspace*{-5mm}}
}}

\maketitle

\begin{abstract}
%\vspace{-2mm}

The industry of wearable remote health monitoring system keeps growing. In the diagnosis of cardiovascular disease, Electrocardiography~(ECG) waveform is one of the major tools which is thus widely taken as the monitoring objective. For the purpose of reducing bit expenditure in the monitoring systems, we study the compression of ECG signal and propose a new compressor in low complexity. Different from the traditional ECG compressors, most of which are built on a single sensor, our compression scheme is based on multiple ECG sensors. The multi-sensor based compression scheme is able to provide more accurate sensing results. Besides the investigation into the structure of the compressor, we also jointly optimize the period and the bit number per sample in the transmission of ECG signal. Experiments are performed on records in MIT-BIH Arrhythmis database and European ST-T database. Experimental results show that our method outperforms conventional ones with respect to ECG reconstruction accuracy at the same bit rate consumption.

\end{abstract}

%\vspace{-5mm}

\section{Introduction}
\label{sec:introduction}

Thanks to the development of mobile communication and positioning technologies~\cite{Lin2017,Huang2014, Huang2011} in the past several decades, remote health monitoring technology is near to practical application in our everyday life. ECG signal is one of the main tools of diagnosing cardiovascular diseases which are the major mortality causes in current societies, especially in developed countries. Remotely monitoring ECG signal provides an effective approach to avoiding the mortality caused by abrupt seizure of cardiovascular diseases.

Basically, in a remote monitoring system a wearable device collects biomedical information, and transmit the collected information to a remote data unit for prompt or delayed diagnosis. The remote monitoring replies on the transmission of the bits which carry ECG signal. The bit transmission induces cost due to consumption of resources provided by infrastructures in mobile communication systems. For the purpose of reducing the cost, we investigate how to reduce the cost with respect to the two aspects, lowering the complexity of ECG compressor and reducing the rate of bits conveying ECG signal. The cost reduction efforts are under the prerequisite that the accuracy of the received ECG signal at the remote data unit should be under control and not degrade the diagnosis of cardiovascular diseases.

\begin{figure}[t!]
\centering
\includegraphics[width=0.8\linewidth] {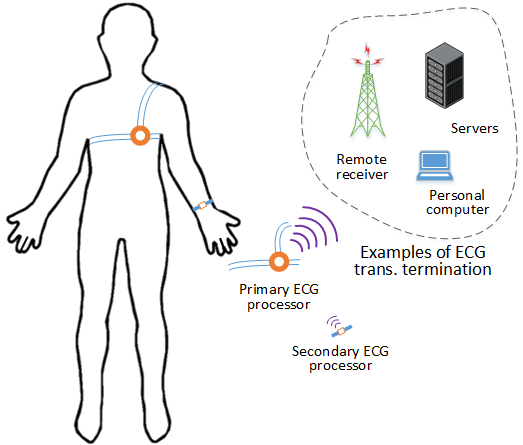}
\setlength{\abovecaptionskip}{3pt plus 3pt minus 2pt}
\caption{Demonstration of two ECG signal compression scheme}
\label{fig:system blocks}
\end{figure}

In literature, a single sensor is widely used to monitor ECG signal. In this paper, we consider multiple-sensor based ECG compression scheme. Generally, a multiple sensors based monitoring system provide more accurate and prompt sensing results since the sensors equipped at different places of our body are able to monitor the conditions of different parts of a heart. Fig.~\ref{fig:system blocks} presents an example of our proposed ECG compressor built on two sensors, a primary ECG sensor and a secondary sensor. As an arbitrary example shown in Fig.~\ref{fig:system blocks}, the secondary sensor is put on the wist which compresses ECG signal and the transmit the compression results to the primary sensor shown in the breast part. Battery capacity and computation capability of the secondary sensor are at a lower than the primary one since the targeted transmission distance of the secondary sensor is shorter. The primary sensor compresses and sends out the ECG signal from the secondary one and itself to a remote data center.

No matter a single- or multiple-sensor based monitoring system, energy consumption is widely recognized as a major concern~\cite{Pandey2011}. The energy consumption is affected by diverse factors, such as hardware chip, circuit board design, encoder, modulation, or even selection of radio frequency~(RF) antenna. Thus, we can hardly evaluate the energy consumption in all terms of the mentioned and unmentioned factors. Independent of the diverse factors, rate of the bits carrying the ECG signal provides us an effective approach to evaluating the energy consumption at a high level.

%%%%%%%%%%%%%%%%%%%%%%%%%%%%%%%%%%%%%%%%%%%%%%%%%%%
%%%%%%%%%%%%%%%%%

Single-sensor based ECG compression scheme have been studied in~\cite{Ma2015,Marisa2015,Deepu2015,Bendifallah2011,Bilgin2003,Lu2000, Zou2015,Polania2015,Cambareri2015, Mamaghanian2011}. Basically, compression methods can be divided into to two categories, direct and differential ECG compression methods.

Uniform quantization is basic one of the direct ECG signal compression methods.  In~\cite{Bendifallah2011}, discrete cosine transform~(DCT) was used to compress ECG signal. Similarly, DCT is also used in ECG signal compression~\cite{Bilgin2003} while Huffman coding was used to compress DCT results further. Still as a direct compression method, wavelet transform followed by run length coding was taken to compress ECG signal in~\cite{Lu2000, Zou2015}. Compressive sensing was utilized to compress ECG signal in~\cite{Cambareri2015, Mamaghanian2011, Polania2015}. 

The compression of ECG signal is implemented in differential structures. Differential schemes built on a linear prediction model are used to compress ECG signal in~\cite{Ruttimann1979, Sun2005}. Multiple ECG samples in the past are taken to predict ECG value in one step ahead. Then, the difference between the prediction and its real value is quantized. In~\cite{Deepu2015,Chen2013,Einarsson1991}, adaptive signal processing methods are taken to update the coefficients of the linear prediction model.

We observe that adjacent ECG samples are not independent to each other. The dependence means there exits redundant information between the samples. Differential compression can effectively reduce the redundant information. After the redundant information reduction, less bits are needed for the quantization. Therefore, differential ECG compression is taken as one of the research objectives in this paper.

There is an important but not solved problem in existing differential ECG compression methods. As we know, coefficients of an adaptive filter for predicting a stationary signal do not change versus time. However, ECG signal is not stationary. Furthermore, ECG signals from a same person can be significantly different. Let us consider such a scenario that one person, sitting on a bench for a long time, stands up to leave. The period of R-R waves in ECG waveform will be different before and after his or her status transition, from sitting to walking. In this case, coefficients of the adaptive filter for predicting his or her ECG waveform will also be different. To keep ECG reconstruction at a high fidelity, the coefficients need to be recalculated and retransmitted; otherwise, there will be huge reconstruction error. To transmit the coefficients of an adaptive filter, a large number of bits will be consumed which is thus harmful to ECG transmission efficiency. In~\cite{Ruttimann1979, Sun2005,Deepu2015,Chen2013,Einarsson1991}, adaptive filters based differential ECG compression schemes are investigated. In their compression schemes, either updating or transmitting coefficients of adaptive filters may cause significant increase of computation resources.

Different from the existing ECG signal compression schemes, we proposed a new structure which is built on multiple ECG sensors. The proposed ECG compressor is at a low complexity. More specifically, the contributions in this paper are presented as follow,

First, we investigate ECG signal compression system with multiple sensors. Simple superposition of multiple sensors is not considered. From a same person, the ECG signals acquired by different ECG sensors at the same time instant usually have similarity in waveform shapes. The similarity means the redundant information. After realizing the signal redundancy between ECG sensors, we design a new ECG compression scheme which effectively saves the bits by reducing the redundancy. 

Second, we propose a novel differential ECG compression scheme which is implemented via comparison and addition operations, and free of multiplications. The traditional differential ECG compressors are built on adaptive filters which rely on the updating filter coefficients and thus increase resource consumptions. This problem does not exists in our compression scheme. Furthermore, we optimize the codebook used for compressing the differential ECG signal.

Third, we optimize compression ECG compression bit rates in two dimensions, the sampling period and the number of bits per sample. To my best knowledge, bit number per sample was considered in literature while the joint optimization is absent.

% With the joint optimization, we are able to effectively save the bit consumption in a new dimension additional to the hardware design optimization.

The remainder of this paper is organized as follows. In Section~\ref{sec: background and system}, the potential problems of the existing compression methods will also analyzed. In Section~\ref{sec: stat of diff ECG}, a novel ECG compression scheme built on multiple sensors will be presented. The joint optimization of bit rate over quantization level and sampling period will be performed in Section~\ref{sec: bit rate opt}. Experiments and simulations will be presented in Section~\ref{sec: experiment} which are followed by conclusions in Section~\ref{sec:conclusion}. 

\section{Related Work and Potential Problems}
\label{sec: background and system}

In this section, we investigate the potential problems in the existing differential ECG compression schemes. Due to the large number of existing reports on ECG compression, our study will not cover all methods but only target at several typical ones.

%  we present differential signal compression methods in literature. Afterwards, problems of the methods potentially existing in ECG compression will be investigated.

\subsection{Open-loop Predictive ECG Compression}
\label{subsec: open loop scheme}

\subsubsection{Open-loop based differential ECG compression method}
\label{subsubsec: procedure of open loop}

\begin{figure}[h]
\centering
\includegraphics[width=0.6\linewidth] {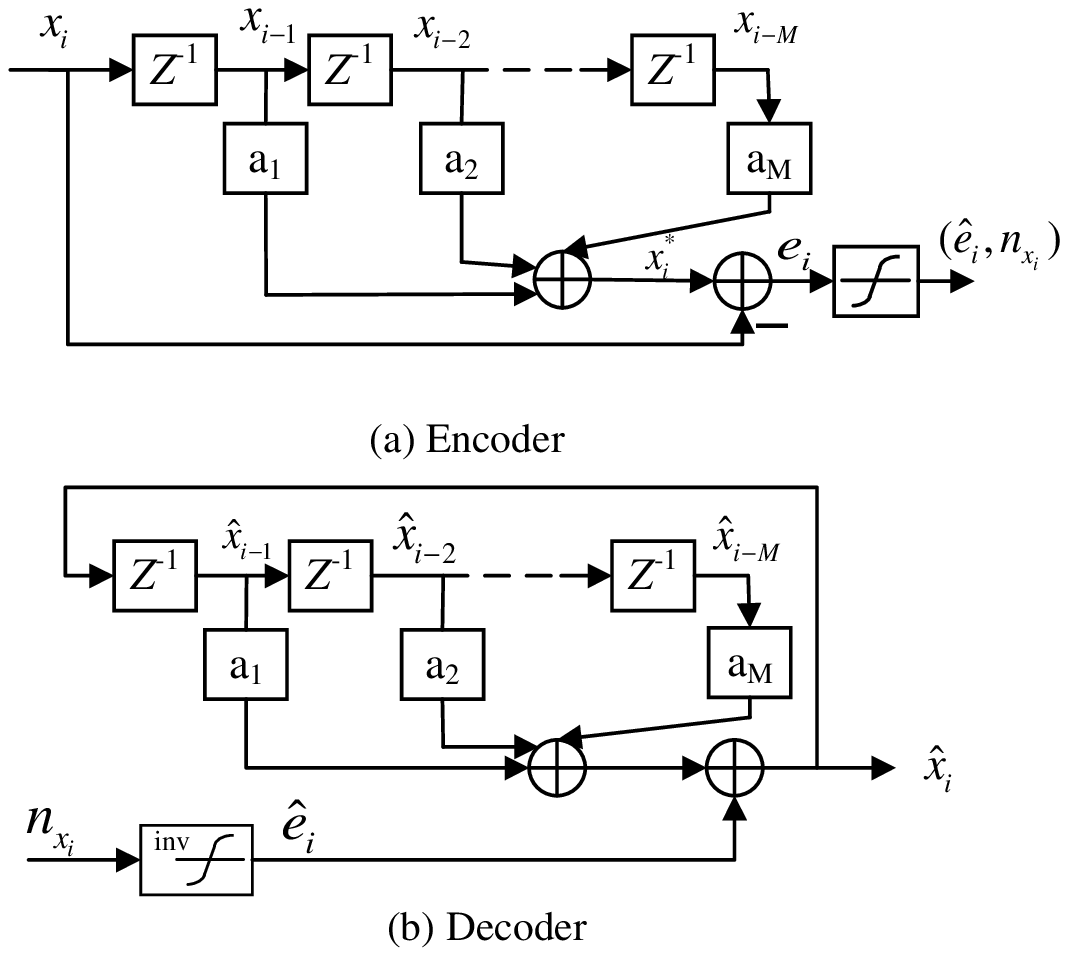}
\setlength{\abovecaptionskip}{3pt plus 3pt minus 2pt}
\caption{Block diagram of open loop differential ECG compressor}
\label{fig:open loop diff ecg comp}
%\vspace{-0.15 in}
\end{figure}

Finite impulse response~(FIR) predicator was widely used in the open-loop based ECG compression. One example of the ECG compressors is shown in Fig.~\ref{fig:open loop diff ecg comp}.

%%%%%%%%%%%%%%%%%%%%%%%%%%%%%%%%%%%%%%%%%%%%%%%%%%%
%%%%%%%%%%%%%%%%%

%system. With the FIR system, a ECG sample at the current time instance is estimated based on samples at past several time instances. 

Let $x(t)$ denote the time continuous ECG signal to be compressed and $x_i$ denote periodical samples of $x(t)$, $i\in\mathbb{Z}$. Assume the FIR predicator is in the order of $M$. Let $a_m$, $m\in\{1,2,\cdots,M\}$ denote coefficients of the predicator. At the $i$-th time instance, estimation of ECG signal is denoted by $x^*_i$ which is calculated as follows
%%%%%%%%%%%%%%%%%%
\begin{equation}
x^*_i=\sum_{m=1}^M a_m x_{i-m}.
\label{eq: fir est fir1}
\end{equation}
%%%%%%%%%%%%%%%%%%

Estimation error between  $x^*_i$ and  $x_i$ is determined by
%%%%%%%%%%%%%%%%%%
\begin{equation}
e_i=x^*_i-x_i.
\label{eq: est err fir1}
\end{equation}
%%%%%%%%%%%%%%%%%%

In a differential ECG signal compression scheme, the estimation error $e_i$ should be encoded and transmitted to a remote receiver. The receiver decodes the codewords and obtains the reconstruction of $e_i$, which is denoted by $\hat{e}_i$. With $\hat{e}_i$, ECG signal is reconstructed by
%%%%%%%%%%%%%%%%%%
\begin{equation}
\hat{x}_i=\hat{e}_i+\sum_{m=1}^M a_m \hat{x}_{i-m}.
\label{eq: recon fir1}
\end{equation}
%%%%%%%%%%%%%%%%%%

The major concern for the open-loop based differential ECG compressor is the stability at the decoder. If errors in quantizing $e_i$ will be accumulated, the compressor system is unstable. Unfortunately, there was no attention paid to the stability problem for open-loop compressors.

\subsubsection{Stability of open-loop based differential ECG compression}
\label{subsubsec: s of open loop}

An unstable open-loop compressor will accumulate quantization errors which will eventually cause the failure of ECG signal reconstruction at the decoder. Therefore, we need to analyze the quantization error accumulation problem at the decoder side. As defined in Section~\ref{subsubsec: procedure of open loop}, $e_i$ is the difference between $x_i$ and its estimation $x^*_i$. At the decoder side, the difference between $x_i$ and its reconstruction $\hat{x}_i$ is denoted by $e_i^*$, 
%%%%%%%%%%%%%%%%%%
\begin{equation}
e^*_i =x_i-\hat{x}_i.
\label{eq: recon err fir1}
\end{equation}
%%%%%%%%%%%%%%%%%%

Furthermore, we define $e_{qi}$ as the difference between $e_i$ and $\hat{e}_i$,
%%%%%%%%%%%%%%%%%%
\begin{equation}
e_i =\hat{e}_i+e_{qi},
\label{eq: quantization err fir1}
\end{equation}
%%%%%%%%%%%%%%%%%%
where $e_{qi}$ is essentially the quantization error in compressing $e_i$. 

We can realize that $e_i^*$ measures the bias of the reconstructed ECG sample with respect to its real value. Only if $e_i^*$ stays within a small bounded range, the decoder is able to obtain accurate ECG samples. The quantization error $e_{qi}$ is the factor which may cause $e_i^*$ to be outside of the bounded range. Therefore, we construct $e_i^*$ as a function of $e_{qi}$. Via analyzing the stability of the function, we can understand whether the ECG compressor is stable. The function is derived as follows,

%%%%%%%%%%%%%%%%%%
\begin{equation}
\begin{aligned}
e^*_i&\overset{(a)}{=}x_i-\left(\sum_{m=1}^M\hat{x}_{i-m}a_m+\hat{e}_i\right)\\
&\overset{(b)}{=}x_i-\left(\sum_{m=1}^M\hat{x}_{i-m}a_m+e_i-e_{qi}\right)\\
&\overset{(c)}{=}x_i-\left(\sum_{m=1}^M\hat{x}_{i-m}a_m+x_i-x^*_i-e_{qi}\right)\\
&\overset{(d)}{=}\sum_{m=1}^M\left(x_{i-m}-\hat{x}_{i-m}\right)a_m+e_{qi}\\
&\overset{(e)}{=}\sum_{m=1}^M e^*_{i-m}a_m+e_{qi},
\end{aligned}
\label{eq: stability e fir1}
\end{equation}
%%%%%%%%%%%%%%%%%%
where $(a)$ follows (\ref{eq: recon fir1}); $(b)$ follows (\ref{eq: quantization err fir1}); $(c)$ follows (\ref{eq: est err fir1}); $(d)$ follows (\ref{eq: fir est fir1}); and $(e)$ follows (\ref{eq: recon err fir1}).

We calculate $z$ transform of (\ref{eq: stability e fir1}) as follows,
%%%%%%%%%%%%%%%%%%
\begin{equation}
%\begin{aligned}
H_{OLP}=\frac{\mathbf{Z}\{e^*\}}{\mathbf{Z}\{e_q\}}=\frac{1}{1-\sum_{m=1}^{M}a_mz^{-m}},
%\end{aligned}
\label{eq: Z trans e fir1}
\end{equation}
%%%%%%%%%%%%%%%%%%
where $\mathbf{Z}\{\cdot\}$ denotes the operator of $Z$ transformation.

The stability of (\ref{eq: Z trans e fir1}) depends on coefficients $a_m$, $m\in\mathcal{M}$. Indeed, shapes of ECG waveforms will differ with different people or different health conditions. The change of ECG waveform generates the different $a_m$. Furthermore, the inconsistence of $a_m$ means no guarantee of the stability in (\ref{eq: Z trans e fir1}). 

%%%%%%%%%%%%%%%%%%%%%%%%%%%%
%\begin{figure}[h]
%\centering
%\includegraphics[width=0.6\linewidth] {Fig/ecg_demo_new}
%\setlength{\abovecaptionskip}{3pt plus 3pt minus 2pt}
%\caption{Examples of ECG waveform for coefficients calculation}
%\label{fig:ECG demo}
%%\vspace{-0.15 in}
%\end{figure}
%%%%%%%%%%%%%%%%%

%Fig.~\ref{fig:ECG demo} $(a)$ and $(b)$ present No. 106 and No.118 records of the MIT-BIH database. 

Aligning with the work in literature, we consider 4-th order FIR predictor. Under MMSE rule, the two sets of $\{a_m\}$ corresponding to No. 106 and No.118 ECG records are equal to $\{ -0.1436, -0.2120, 0.1582, 1.1548\}$ and $\{-0.2276, -0.2041, 0.2512, 1.1761\}$ respectively. With the calculated coefficients, $H_{OLP}|_a$ and $H_{OLP}|_b$, are correspondingly determined. Then, poles of the two impulse response functions are calculated which are equal to $p_a=\{-0.9823, -0.0761\pm j1.0866,  0.9908\}$ and $p_a=\{-0.9868,-0.1200 \pm j1.0856,  0.9991\}$ respectively. From the poles, we can easily realize that $H_{OLP}|_a$ and $H_{OLP}|_b$ are not necessary to be stable which means there exists the risk of inducing the failure of ECG reconstruction at the decoder.  

% are utilized to calculate the coefficients. The two ECG records are plotted in  respectively. 

 %If coefficients of FIR system are not updated, stability of FIR based differential ECG compressor can hardly be guaranteed. If the coefficient are updated, there is another problem that transmission efficiency will be low. As introduced, 

\subsection{Closed-Loop Predictive ECG Compression}
\label{subsec: closed loop scheme}

%\subsection{Open-loop Predictive ECG Compression}
%\label{subsec: open loop scheme}

From Section~\ref{subsubsec: s of open loop}, open-loop differential ECG compressors have the risk of being instable at the decoder. This problem can be solved by adding a feedback to the quantization of $e_i$.

\subsubsection{Closed-loop differential ECG compression method}
\label{subsubsec: procedure of closed loop}

 The differential compressor with a feedback is called as closed-loop differential ECG compressor. Still $M$ denotes the order of the linear model used to estimate the value of an ECG sample. When $M=1$, the differential compressor degenerates into Differential pulse code modulation~(DPCM).  

%%%%%%%%%%%%%%%%%%%%%%%%%%%%%%%%%%%%%%%%%%%%%%%%%%%
\begin{figure}[h]
\centering
\includegraphics[width=0.6\linewidth] {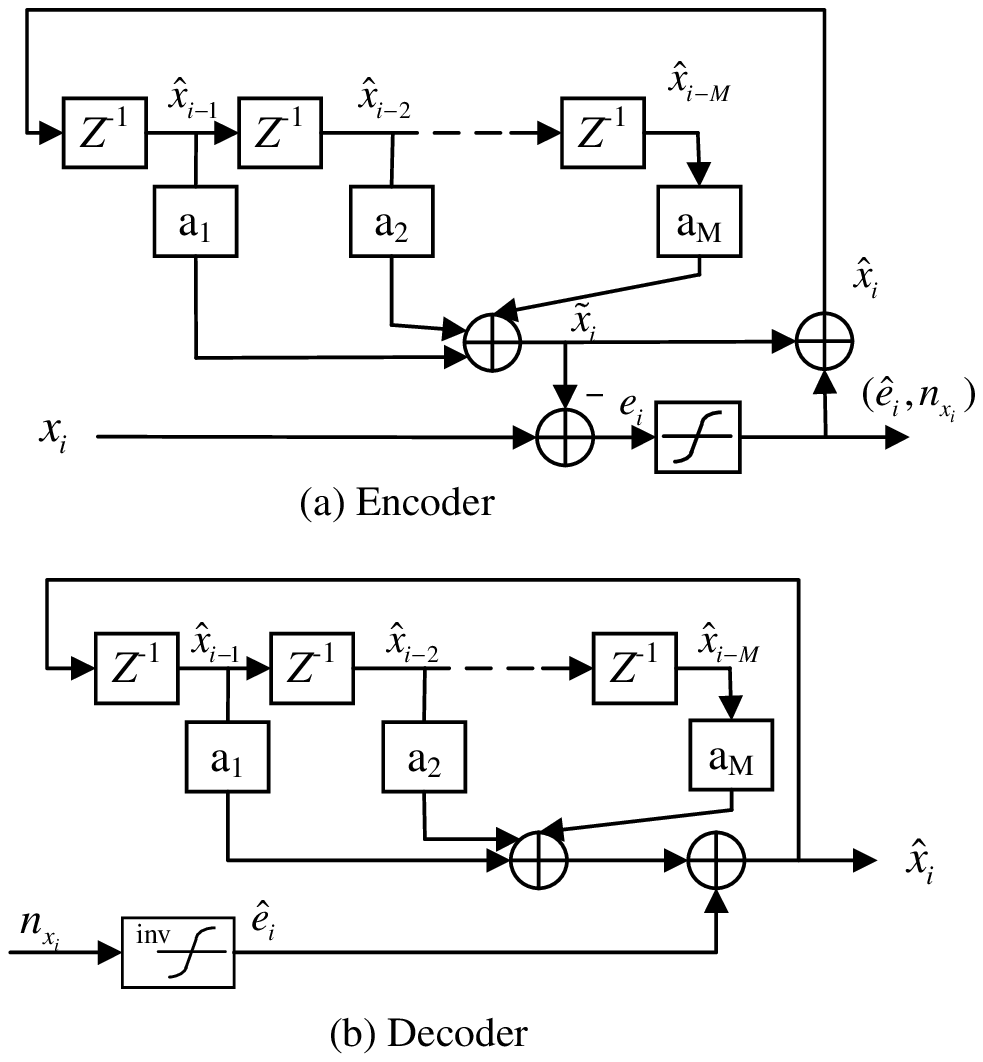}
\setlength{\abovecaptionskip}{3pt plus 3pt minus 2pt}
\caption{Block diagram of closed loop differential ECG compressor}
\label{fig:closed loop diff ecg comp}
%\vspace{-0.15 in}
\end{figure}
%%%%%%%%%%%%%%%%%

Fig.~\ref{fig:closed loop diff ecg comp} plots the block diagram of closed-loop differential compressors. Compared with open-loop compressor, the major difference in the closed-loop one is that reconstruction is performed at the encoder side, and the reconstructed sample is taken as a reference of modifying the threshold for quantizing the next ECG sample. 

Let $\tilde{x}_i$ denote the estimation of ECG sample at $i$-th time instance at the encoder side which is calculated by
%%%%%%%%%%%%%%%%%%
\begin{equation}
\tilde{x}_i=\sum_{i=m}^Ma_m\hat{x}_{i-m}.
\label{eq: predict DS}
\end{equation}
%%%%%%%%%%%%%%%%%%

The estimation bias $e_i$ is determined as follows,
%Then, difference between original current value and its estimation is calculated via 
%%%%%%%%%%%%%%%%%%
\begin{equation}
e_i=x_i-\tilde{x}_i.
\label{eq: e cal DS}
\end{equation}
%%%%%%%%%%%%%%%%%%

Afterwards, $e_i$ is first quantized and the quantization result is denoted by $\hat{e}_i$, and the quantization error is stilled represented by $e_{qi}$. At the encoder side, the reconstruction of an ECG sample, denoted by $\hat{x}_i$, is obtained by adding the quantized $e_i$ to $\tilde{x}_i$,
%%%%%%%%%%%%%%%%%%
\begin{equation}
\hat{x}_i=\tilde{x}_i+\hat{e}_i.
\label{eq: feedback DS}
\end{equation}
%%%%%%%%%%%%%%%%%%

As shown in (\ref{eq: feedback DS}), $\hat{x}_i$ is feed back to the input of the linear filter. Since $\tilde{x}_i$ contains the error occurring in the quantization of the previous ECG sample, the feedback is beneficial for avoiding the accumulation of the quantization error. 

\subsubsection{Stability of closed-loop based differential ECG compression}
\label{subsubsec: stability of closed loop}

Let $e^*_{Ci}$ denote the difference between the ECG sample $x_i$ and its reconstruction at the decoder. For closed-loop compressor, the reconstructions of an ECG sample at both the encoder and decoder are the same. Therefore, the reconstruction at the decoder is also denoted by $\hat{x}_i$.  Due to the same reason mentioned in Section~\ref{subsubsec: s of open loop}, we calculate $e^*_{Ci}$ as a function of $e_{qi}$. Via analyze the stability of the calculated function, we can understand whether there exists the risk of accumulating quantization errors. The calculation of the function $e^*_{Ci}$ of $e_{qi}$ is presented as follows,

%%%%%%%%%%%%%%%%%%
\begin{equation}
\begin{aligned}
e^*_{Ci}&\overset{\Delta}{=}x_i-\hat{x}_i\overset{(a)}{=}x_i-\left(\sum_{i=m}^Ma_m\hat{x}_{i-m}+\hat{e}_{i}\right)\\
&=x_i-\left(\sum_{i=m}^Ma_m\hat{x}_{i-m}+e_{i}-e_{qi}\right)\\
&\overset{(b)}{=}x_i-\left(\sum_{i=m}^Ma_m\hat{x}_{i-m}+x_i-\sum_{i=m}^Ma_m\hat{x}_{i-m}-e_{qi}\right)\\
&=e_{qi}\\
\end{aligned}
\label{eq: cal of e* Delta}
\end{equation}
%%%%%%%%%%%%%%%%%%
where $(a)$ follows (\ref{eq: feedback DS}); $(b)$ follows (\ref{eq: predict DS}) and (\ref{eq: e cal DS}).

From (\ref{eq: cal of e* Delta}), ECG reconstruction error in the closed-loop compressor is fully determined by the error in quantizing $e_i$. In practice, quantization error is finite in a given quantizer. Therefore, the closed loop ECG compressor is always stable. 

\section{Proposed ECG Compression Scheme Based on Multiple Sensors}
\label{sec: stat of diff ECG}

Besides the absence of the stability analysis of ECG compressor, there is another unsolved problem in the existing studies, that is, only signal sensor is considered to compress ECG signal. Indeed, more sensors are able to provide more observations on the heart conditions since ECG signals obtained by sensors placed on different places of a body reflect the health conditions of different parts of a heart. Therefore, we investigate the ECG signal compression based on multiple sensors.  

For multiple sensors, independent quantization is an inefficient practice since the redundancy between ECG signals from the multiple sensors is not removed. The retaining redundant information induces more bits for quantization. We propose a compression method used for multiple sensors. 

The multiple sensors are divided into two tiers, that is, one primary sensor is taken as the first tier and the all the other sensors are at the secondary tier. The primary sensor has more powerful computation and transmission abilities which is responsible for remotely transmitting the ECG signal. The secondary sensors transmit their collected ECG signal to a primary one and the transmission range is smaller than that for the primary sensor. At the secondary sensor, conditional quantizer is used to compress ECG signal which can effectively reduce the redundant information. For analysis simplicity, we consider the case with one primary sensor and one secondary sensor.

\subsection{Structure of Multiple Sensors Based ECG Compression Scheme}
\label{subsec: structure of new}

\subsubsection{System Overview}
\label{subsubsec: system}

%In a symmetric structure, two sensors have the same complexity and battery such that they independently compress and transmit ECG signal to remote data unit. This symmetric structure is not economical.

\begin{figure}[h]
\centering
\includegraphics[width=0.8\linewidth] {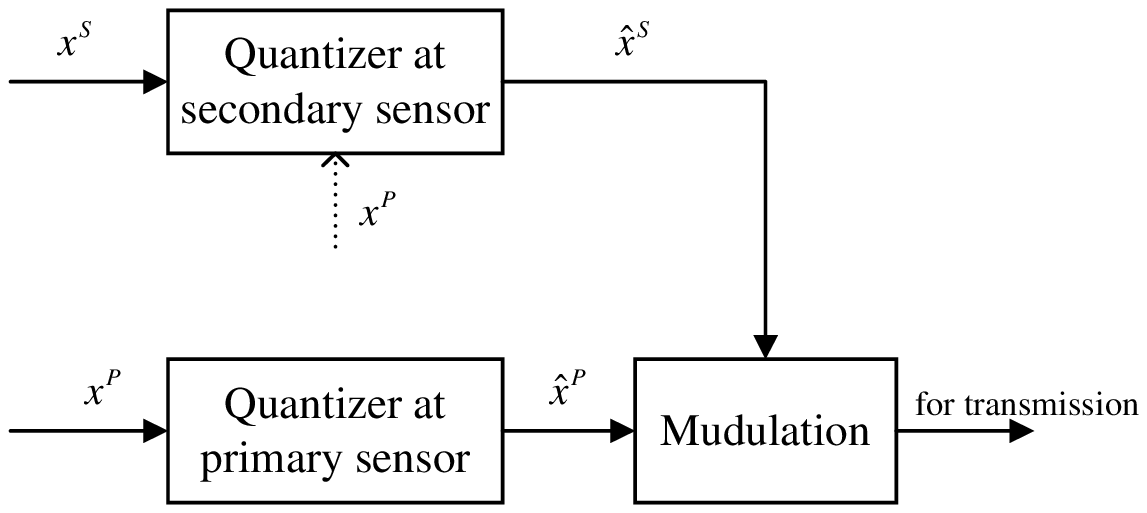}
\setlength{\abovecaptionskip}{3pt plus 3pt minus 2pt}
\caption{Block diagram of double sensor quantization scheme}
\label{fig: system}
%\vspace{-0.15 in}
\end{figure}

Fig.~\ref{fig: system} presents the block diagram of the compression scheme built on the primary sensor and secondary sensor. The secondary sensor transmits quantized ECG signal $\hat{x}^S$ to the primary one. The primary sensor quantizes $x^P$ to obtain $\hat{x}^P$ and transmits the two quantized ECG signals ($\hat{x}^S$ and $\hat{x}^P$) to a remote data unit. In the scheme, waveform features of $\hat{x}^P$ are priorly known by the secondary sensor.

%%%%%%%%%%%%%%%%%
%%%%%%%%%%%%%%%%%

With the waveform feature of $\hat{x}^P$, we perform conditional quantization at the secondary sensor. The output from the conditional quantizer at the secondary sensor is sent to the primary one. The primary sensor takes differential compression scheme to quantize ECG signal. 

After introducing the functions of the modules in the compression scheme, we present the details of how to implement the differential compression method at the primary sensor which is followed by the stability analysis. Then, the conditional quantization at the secondary sensor is introduced.

\subsubsection{Differential compression scheme at primary ECG sensor}
\label{subsubsec: single structure}

\begin{figure}[h!]
\centering
\includegraphics[width=0.7\linewidth] {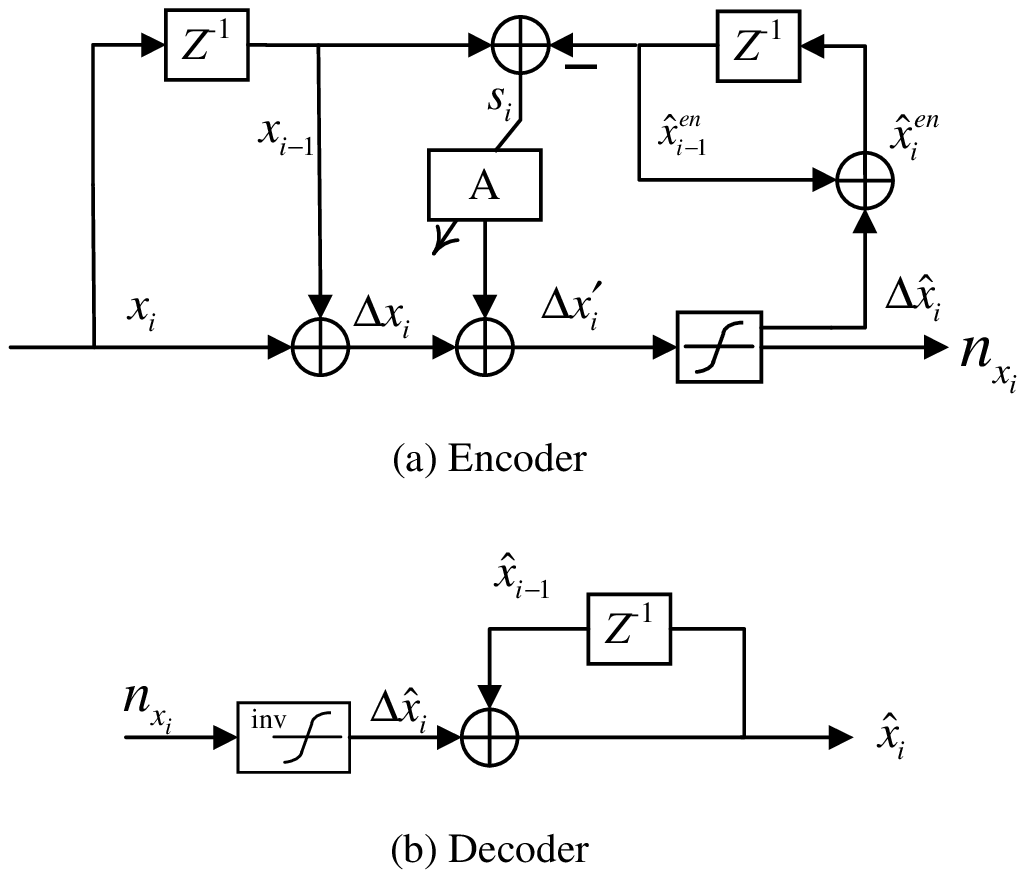}
\setlength{\abovecaptionskip}{3pt plus 3pt minus 2pt}
\caption{Block digram of proposed ECG data compression scheme}
\label{fig:proposed system}
\end{figure}

Block diagram of our proposed differential ECG compressor is presented in Fig.~\ref{fig:proposed system}. Fig.~\ref{fig:proposed system} (a) and (b) describe the encoder and decoder respectively. Compared with conventional closed-loop compressors, only addition and comparison operations are needed, and multiplication is absent in the proposed one. Furthermore, we will illustrate our compressor outperforms the conventional ones in terms of ECG reconstruction accuracy.

%%%%%%%%%%%%%%%%%%%%%%%%%%%%%%%%%%%%%%%%%%%%%%%%%%%
%%%%%%%%%%%%%%%%%

In the new differential compressor, the difference between two adjacent ECG samples is first calculated as follows, 
%%%%%%%%%%%%%%%%%%
\begin{equation}
\Delta x_i=x_i-x_{i-1}.
\label{eq: Delta x cal Huang}
\end{equation}
%%%%%%%%%%%%%%%%%%

Next, a modification factor, denoted by $A$, is added to $\Delta x_i$. The factor $A$ is designed to counteract the accumulation of quantization errors. After the addition of $A$, the adjacent difference $\Delta x_i$ is derived into $\Delta x^{\prime}_i$ as follows,
%%%%%%%%%%%%%%%%%%
\begin{equation}
\Delta x^{\prime}_i=\Delta x_i+A.
\label{eq: Delta x modification Huang}
\end{equation}
%%%%%%%%%%%%%%%%%%

Afterwards, $\Delta x^{\prime}_i$ is quantized and quantization index is denoted by $l_{i}$. The quantization levels constitute a set $\Phi$, $\Phi=\{\phi_l, l\in\mathcal{L}\}$, $\mathcal{L}=\{1,2,3,\cdots,L\}$, where $N$ is equal to the total number of quantization levels. The elements in $\Phi$ are ordered incrementally. As we know, the set $\Phi$ has impact on the quantization performance. The details in designing $\Phi$ will be introduced in Section~\ref{subsubsec: quantization primary}. After the quantizing $\Delta x^{\prime}_i$, the compressor will perform modulation according to $l_i$. 

%will be introduced in following sections. Then, 

From the previous paragraph, the modification factor $A$ is an important parameter. Next, we discuss the details of how to determine $A$. 

Let $\Delta\hat{x}_i$ denote the quantized $\Delta x^{\prime}_i$. To reconstruct the ECG sample at $i$-th time instance ($\hat{x}_{i}$) at the encoder, we add the $\Delta\hat{x}_i$ to $\hat{x}_{i-1}$,
%.  which is the reconstructed ECG at previous one time instance. Thus, we obtain the reconstructed ECG at current time instance, $\hat{x}_{i}$, where
%%%%%%%%%%%%%%%%%%
\begin{equation}
\hat{x}_{i}=\hat{x}_{i-1}+\Delta \hat{x}_i.
\label{eq: recon of x encoder}
\end{equation}
%%%%%%%%%%%%%%%%%%

Since $A$ is used to counteract the accumulation of quantization error, $A$ is designed to be a function for reducing the quantization error at the previous time instance. The quantization error at the previous time instance is calculated by 
%%%%%%%%%%%%%%%%%%
\begin{equation}
s_{i}=x_{i-1}-\hat{x}_{i-1}.
\label{eq: compare flag}
\end{equation}
%%%%%%%%%%%%%%%%%%

To avoid the accumulation of quantization error, $A$ is set to be a small positive value, when $s_{i}<0$; and $A$ is a negative value, when $s_{i}\geq0$. This correlation between $A$ and $s_{i}<0$ is mathematically described by 
%%%%%%%%%%%%%%%%%%
\begin{equation}
A=\left\{\begin{matrix}
\left(\phi_{l_{\Delta x^{\prime}_i}}-\phi_{l_{\Delta x^{\prime}_i}-1}\right) & \mbox{if} & s_i>0\\
-\left(\phi_{l_{\Delta x^{\prime}_i}+1}-\phi_{l_{\Delta x^{\prime}_i}}\right) & \mbox{if} & s_i<0
\end{matrix}\right.,
\label{eq: A determine}
\end{equation}
%%%%%%%%%%%%%%%%%%
where $l_{\Delta x^{\prime}_i}$ denotes the index of the quantized $\Delta x^{\prime}_i$ within the ordered set $\Phi$, and $\phi_{l_{\Delta x^{\prime}_i}}$ is the corresponding quantization result.

\textbf{Stability Analysis}: To analyze the stability of the proposed compression scheme, we derive the compression bias $e^*_i$ as a function of the quantization error $e_{qi}$ due to the same reason presented in Section~\ref{subsubsec: s of open loop}. The quantization error $e_{qi}$ satisfies the following equation,
%%%%%%%%%%%%%%%%%%
\begin{equation}
e^{\prime}_i=\hat{e}_i+e_{qi}.
\label{eq: quan error def huang}
\end{equation}
%%%%%%%%%%%%%%%%%%

For analysis convenience, we simplify (\ref{eq: A determine}) into a form as follows,

% Similar with (\ref{eq: recon err fir1}), we also define the error in reconstructing ECG signal which is denote by $e^*_i$. 
%%%%%%%%%%%%%%%%%%
\begin{equation}
A=\beta\left(x_{i-1}-\hat{x}_{i-1}\right),
\label{eq: A simplify}
\end{equation}
%%%%%%%%%%%%%%%%%%
where $\beta$ is variable which absolute value is bounded into a small range and the sign of $\beta$ is opposite to the sign of $(x_{i-1}-\hat{x}_{i-1})$. 

%Next, we will show our proposed ECG compressor is stable which thus will not accumulate errors. 

Next, we determine the expression of $e_i^*$ as follows,
%%%%%%%%%%%%%%%%%%
\begin{equation}
\begin{aligned}
e^*_i&=x_i-\hat{x}_i=x_i-\left(\Delta\hat{x}_i+\hat{x}_{i-1}\right)\\
&\overset{(a)}{=}x_i-\left(\Delta x_i+A-e_{qi}+\hat{x}_{i-1}\right)\\
&\overset{(b)}{=}x_i-\left(x_i-x_{i-1}+A-e_{qi}+\hat{x}_{i-1}\right)\\
&\overset{(c)}{=}\left(x_{i-1}-\hat{x}_{i-1}\right)-\left(\beta\left(x_{i-1}-\hat{x}_{i-1}\right)-e_{qi}\right)\\
&=\left\{\begin{matrix}
 \left(1-|\beta|\right)e^*_{i-1}+e_{qi}& \mbox{for}&x_{i-1}-\hat{x}_{i-1}>0\\
 \left(1+|\beta|\right)e^*_{i-1}+e_{qi}& \mbox{for}&x_{i-1}-\hat{x}_{i-1}<0\\
 \end{matrix}\right.\\
&=\left(1-|\beta|\right)e^*_{i-1}+e_{qi}.
\end{aligned}
\label{eq: stability e huang}
\end{equation}
%%%%%%%%%%%%%%%%%%
where $(a)$ follows (\ref{eq: quan error def huang}) and (\ref{eq: Delta x modification Huang}); $(b)$ follows (\ref{eq: Delta x cal Huang}); $(c)$ follows (\ref{eq: A simplify}).

The $Z$-transformation of (\ref{eq: stability e huang}) is written as
%%%%%%%%%%%%%%%%%%
\begin{equation}
\begin{aligned}
H(z)=\frac{\mathbf{Z}\{e^*\}}{\mathbf{Z}\{e_q\}}=\frac{1}{1-(1-|\beta|)z^{-1}}.
\end{aligned}
\label{eq: Z trans e huang}
\end{equation}
%%%%%%%%%%%%%%%%%%

From (\ref{eq: Z trans e huang}), the pole is equal to $p=1-|\beta|$ which locates in inner of a unit circle. Therefore, our proposed ECG processor can avoid the accumulation of quantization error.

%Compare with DPCM based ECG compressor, the proposed one attenuates quantization error at the rate of $(1-|\beta|)$, while conventional closed-loop differential compressor does not. Therefore, the proposed method is able to achieve more accurate ECG reconstruction. 

% Thus, resolution of the quantizer fully determines the reconstruction error. 

\subsection{Quantizer Design in the Proposed ECG Compressor}
\label{subsec: quantization of diff ecg}

In the optimum sense of minimizing average quantization error at a given number of quantization levels, the statistics of the quantization objective affects the design of an optimum quantizer. Thus, we first analyze the statistic features of the differential ECG signal. Afterwards, we present the details of how to design the differential ECG compressor. Then, the conditional quantization by the secondary sensor is introduced.

\subsubsection{Statistical Features of One-step Differential ECG Data} 
\label{subsubsec: dynamic range}

There are two important issues determining statistical features of a signal, dynamic range of source and distribution of it. We will numerically analyze the differential ECG signal at the two aspects using  two factors

%. Since ECG signal is not stationary, we prefer  methods to analyze statistical feature of ECG signal. 

\begin{figure}[h]
\centering
\includegraphics[width=0.81\linewidth] {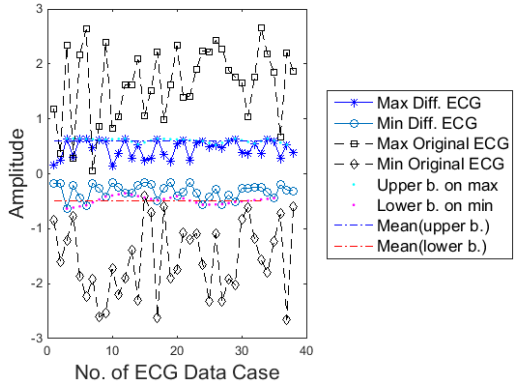}
\setlength{\abovecaptionskip}{3pt plus 3pt minus 2pt}
\caption{Dynamical range of differential ECG and original ECG signal}
\label{fig:dynamic range}
%\vspace{-0.15 in}
\end{figure}

Fig.~\ref{fig:dynamic range} shows us the dynamical ranges of differential ECG and original ECG waveform which are calculated from 38 records in MIT-BIH database. 

%%%%%%%%%%%%%%%%%%%%%%%%%%%%%%%%%%%%%%%%%%%%%%%%%%%
%%%%%%%%%%%%%%%%

For each record, we calculate the maximum and minimum values of both original and differential ECG signals. All the extreme values are plotted in Fig.~\ref{fig:dynamic range}. 

To determine the dynamical range of differential ECG signal, we first calculate the upper bound of the maximum points and lower bound of minimum points via linear interpolation. Afterwards, we perform curve fitting on the two bounds using two horizontal lines. The two horizontal lines label the boundaries of the differential ECG dynamical range. The same method is also used to determine the dynamical range of the original ECG signal. From the calculation, we can observe that dynamical range of original ECG signal is approximately equal to 6. The differential ECG data has the dynamic range from -0.4854 to 0.6044. Since the dynamical range of the differential ECG signal is smaller than that of the original signal, less bits are needed for quantizing the differential ECG signal at a given quantization accuracy. 

%, we plot maximum and minimum values of original and differential ECG signal of each record. 

\begin{figure}[h]
\centering
\includegraphics[width=1\linewidth] {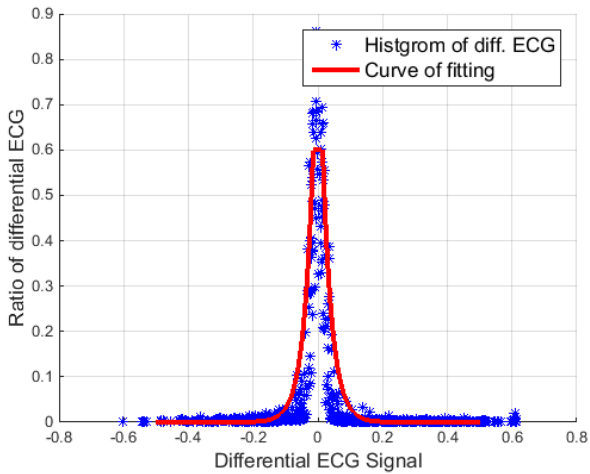}
\setlength{\abovecaptionskip}{3pt plus 3pt minus 2pt}
\caption{Approximation of histogram via curve fitting using exponential functions}
\label{fig: dist approx}
%\vspace{-0.15 in}
\end{figure}

After analyzing the dynamical range, we study the distribution of the differential ECG signal. First, we calculate histogram of differential ECG signal which is plotted by the blue stars in Fig.~\ref{fig: dist approx}. With the calculated histogram, we use the curve fitting technology to abstract an approximated probability model of the differential ECG signal.

%%%%%%%%%%%%%%%%%%%%%%%%%%%%%%%%%%%%%%%%%%%%%%%%%%%
%%%%%%%%%%%%%%%%

Let $f(\Delta x)$ denote the probability. An exponential function with peak clipping is used to represent the differential ECG histogram. The red bold curve in Fig.~\ref{fig: dist approx} plots the probability function $f(\Delta x)$ which takes a form of exponential function with the exponent of -50. With the calculated probability model, we analytically study the quantizer design in the following part. 

%, and the function is plotted by the bold red curve. 

%\subsubsection{Statistical Features of Differential ECG Signals over Two Sensors} 
%\label{subsubsec: stat feature of 2 sensor}

\subsubsection{Quantization of differential ECG at primary sensor} 
\label{subsubsec: quantization primary}

Assume the quantizer in our analysis is labeled by $Q$. The quantizer is optimized in the sense of minimum mean square of quantization error. The mean square error is calculated by
%%%%%%%%%%%%%%%%%
\begin{equation}
esq_Q=\sum_{l=0}^{L-1}\int_{\Delta x_l}^{\Delta x_{l+1}}\left(\Delta x-\Delta x^0_l\right)^2 f(\Delta x) d\Delta x,
\label{eq: def quan mse }
\end{equation}
%%%%%%%%%%%%%%%%%
where $L$ is the number of total quantization levels and $x^0_l$ is quantization output at $l$-th quantization level. 

We select Lloyd-Max algorithm~\cite{Lloyd1982} to determine each quantization zone $(\Delta x_l, \Delta x_{l+1})$ and the value of quantization output $\Delta x^0_l$. According to Lloyd-Max algorithm, the parameters are iteratively calculated as follows 
%%%%%%%%%%%%%%%%%
\begin{equation}
\Delta x^0_l=\frac{\int_{\Delta x_l}^{\Delta x_{l+1}}\Delta x f(\Delta x)d\Delta x}{\int_{\Delta x_l}^{\Delta x_{l+1}} f(\Delta x)d\Delta x},
\label{eq: quantization output}
\end{equation}
%%%%%%%%%%%%%%%%%
%%%%%%%%%%%%%%%%%
\begin{equation}
\Delta x_l=\frac{\Delta x^0_l+\Delta x^0_{l+1}}{2}.
\label{eq: int zone b}
\end{equation}
%%%%%%%%%%%%%%%%%

In a partial summarization of the quantization on the primary sensor, the histogram of the first order differential ECG signal is calculated first; second, via curve fitting, a PDF in an explicit form is calculate to approximate the histogram; third, the number of the bits for the quantization is determined; fourth, the codebook and quantization zones are determined according to (\ref{eq: quantization output}) and (\ref{eq: int zone b}) respectively.

\subsubsection{Quantization on the Secondary ECG Sensor} 
\label{subsubsec: quantization secondary}

\begin{figure}[h]
\centering
\includegraphics[width=0.6\linewidth] {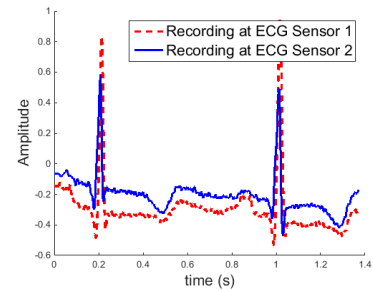}
\setlength{\abovecaptionskip}{3pt plus 3pt minus 2pt}
\caption{Number 100 ECG recording in MIT-BIH Arrhythmia database}
\label{fig: MIT ECG 100}
%\vspace{-0.15 in}
\end{figure}

The proposed differential ECG compression method at the primary sensor achieves the bit rate saving by reducing the redundancy between ECG samples from a same sensor. Besides, the redundancy within the ECG samples from a single sensor, there also exits inter-sensor redundancy which can be observed from the waveform similarities between the ECG signals from different sensors. Without loss of generality, No. 100 ECG recording in MIT-BIH arrhythmia database is plotted in Fig.~\ref{fig: MIT ECG 100} which is taken as an example of showing the existence of inter-sensor redundancy. We will reduce to the inter-sensor redundancy to save the bit rate for the quantization on the secondary sensor.

%shows No. 100 ECG recording  From Fig.~\ref{fig: MIT ECG 100}, we can easily observe the similarity between signals obtained by two ECG senors. The similarity reflects the redundant information between the two sensors. 

%The differential compression method reduces bit rate via utilizing the redundant information between adjacent ECG sample.

%The dependency between the primary and secondary ECG sensors can be taken as the measurement of the redundant information. 

%%%%%%%%%%%%%%%%%%%%%%%%%%%%%%%%%%%%%%%%%%%%%%%%%%%
%%%%%%%%%%%%%%%%%

%For denotation convenience, the primary sensor is labeled with the number 1. 

%ke sensor 1 as the

The ECG data collected by the primary sensor is denoted by $x^P$, and $x^S$ is for the data from the secondary sensor. Let $f_P(\cdot)$ and $f_S(\cdot)$ denote the approximated PDFs from $x^P$ and $x^S$ respectively. At the secondary sensor, the approximated PDF of $x^P$ is priorly known. The quantization based on the prior information is denoted by $Q(x^S|f_P)$. 

Since there exists the connection between the waveforms of $x^P$ and $x^S$, we build an affine relation between $f_P(\cdot)$ and $f_S(\cdot)$. The affine is denoted by $\digamma$ which is determined as follows,
%%%%%%%%%%%%%%%%%%
\begin{equation}
\begin{aligned}
\digamma=\{(a,b)|\min(f_P-a\cdot f_S-b)^2\},
\end{aligned}
\label{eq: affine def}
\end{equation}
%%%%%%%%%%%%%%%%%%
where $a$ and $b$ are constants for a group of ECG data from a same person, such as the group consisting of $x^P$ and $x^S$.

%The quantization bins can also be determined by Lloyd-Max algorithm. 

With the established affine relation, the conditional quantization and codebook can be calculated according to the following four steps: 
\begin{itemize}
  \item First, a small number of bits, which number is denoted by $N_1$, are used to quantize the support area of $f_S(\cdot)$. Since bits number is smaller, the quantization bins are sparse in the step. Let $b_{1i}^S$ denote the $i$-th quantization bin. Boundaries of $b_{1i}^S$ are $x_{1i}^S$ and $x_{1(i+1)}^S$, where $x_{1i}^S\leq b_{1i}^S<x_{1(i+1)}^S$ and $i\in\{0,1,2,\cdots,2^{N_1}\}$. 
  \item We calculate the boundaries $x_{i}^P$ from $x_{i}^S$ according to the $\digamma$ affine defined in (\ref{eq: affine def}). More explicitly, the calculation is presented below
%%%%%%%%%%%%%%%%%%
\begin{equation}
\begin{aligned}
x_{i}^P=a\cdot x_{1i}^S+b.
\end{aligned}
\label{eq: affine cal}
\end{equation}
%%%%%%%%%%%%%%%%%%
  \item Within each bin of $b_{1i}^S$, $i\in\{0,1,2,\cdots,2^{N_1}\}$, we utilize Lloyd-Max algorithm to calculate a sub-codebook which is denoted by $c_i^{S}$. Let $N_2$ denote the number of bits used in the sub-level quantization.
\item Using the calculated sub-codebooks, we quantize ECG signal within all bins of $b_{1i}^S$, $i\in\{0,1,2,\cdots,2^{N_1}\}$. The corresponding quantization indexes, denoted by $I^s$, are the final outputs of the compressor on the secondary sensor. 
\end{itemize}

\begin{figure}[h]
\centering
\includegraphics[width=0.6\linewidth] {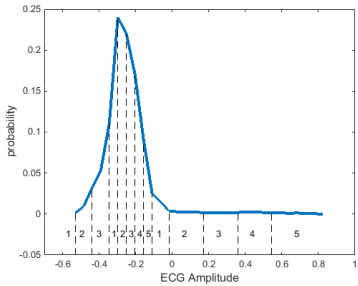}
\setlength{\abovecaptionskip}{3pt plus 3pt minus 2pt}
\caption{Demon of compression on secondary ECG sensor}
\label{fig: secondary comp demon}
%\vspace{-0.15 in}
\end{figure}

To assist our explanation, Fig.~\ref{fig: secondary comp demon} presents a toy example of the conditional quantization method. In Fig.~\ref{fig: secondary comp demon}, the first three segments labeled with 1, 2, and 3 constitute a set. Near to the first set, the 5 numbers ($\{1,2,3,4,5\}$) labeled five segments form the second set. Beside the second one, the third set is constituted in the same way. Each of three sets ($\{1,2,3\}$, $\{1,2,3,4,5\}$ and $\{1,2,3,4,5\}$) covers the range of a $b_{1i}^S$, $i\in\{1,2,3\}$, and all the three sets cover the full dynamic range of $x^S$ without overlapping. In each bin of $b_{1i}^S$, sub-codebook is calculated following the third step above. Then, the quantization is performed in each bin according to the calculated codebook and the numbers noted in Fig.~\ref{fig: secondary comp demon} are the final results of the compression on the secondary sensor. 

%  the  numbers from 1 to 5 are fin The bold blue curve is the estimated PDF of ECG signal at the secondary sensor. The numbers placed among adjacent dash lines represent the indexes of the quantization results. 

%Fig. 10 is used to illustrate the basic idea of the compression on the secondary ECG compressor. In Fig. 10, we call the group of numbers {1,2,3,4,5} as a subset. All subsets cover the full range of ECG amplitude at the secondary compressor without overlapping. At the secondary compressor, the first thing to be determined is which subset an ECG sample locates in. After knowing the subset the sample belonging to, we quantize the sample using five levels. Then, the quantization results are transmitted to the primary compression scheme. It’s worth noting that the subset with the cardinality of 5 is an example for explanation. 

%%%%%%%%%%%%%%%%%%%%%%%%%%%%%%%%%%%%%%%%%%%%%%%%%%%
%%%%%%%%%%%%%%%%%

After the conditional quantization, the secondary sensor sends the quantization results to the primary sensor where the reconstruction is performed. To reconstruct ECG signal at the decoder, two steps are needed.  
\begin{itemize}
  \item Among the set of sections $[x_{i}^P, x_{i+1}^P)$, $0\leq i \leq 2^{N_1}$, we determine which section $x^P$ belongs to. For example, $x^P$ belongs to the $k$-th section, $x_{k}^P\leq x^P < x_{k+1}^P$. 

\item We take the $k$-th sub-codebook $c_k^S$ to determine the reconstruction corresponding to the quantization index $I^s$.
\end{itemize}

Based on the description above, we can realize that there are $N_2$ bits used in the compression at the secondary sensor. In the experiments of this paper, we will show that $N_2$ is smaller than the number of bits for direction quantization, $N_2<W$, where $W$ denote the number of the bits used for direct quantization. 

In this section, the structure of implementing the proposed ECG compression scheme is presented which is at low complexity and thus easy to be implemented at less hardware resource cost. Furthermore, the new compression scheme saves the consumed bits per sample. Indeed, besides the bits per sample, the sampling rate will also affect the accuracy of ECG compression and the hardware resource consumption. In the next section, we perform joint optimization with respect to the two aspects.

\section{Two Dimensional Bit Rate Optimization}
\label{sec: bit rate opt}

% The strategy of reducing quantization bits on a low complexity structure is analyze in the previous several sections. In this section, we jointly optimize the bit number and transmission period. 

In a remote health monitoring system, there are two factors significantly affecting the complexity and power consumption, the quantization bits per sample and and transmission period. Let $W$ denote the average number of the bits used for quantizing each sample, and $T_t$ is the period of transmitting the compressed ECG data. The bit rate $r$ is calculated by $r=\frac{W}{T_t}$. In the optimization, our objective is to minimize average square error of the reconstructed ECG signal. The minimization is under the constraint of a given bit rate. With respect to the single dimension of bits number, the related minimization work refers to~\cite{Huang2014a} The theoretic knowledge about the two-dimension optimization refers to~\cite{Huang2014b,Huang2016,Huang2017}.

For an ECG sensor, let $T_0$ denote the minimum sampling period. After the sampling, the ECG signal written as $x(mT_0)$, $m\in\mathbb{Z}$. In practice, the sampling frequency $\frac{1}{T_0}$ is over high for ECG signal. Thus, the ECG signal tor be transmitted should be down sampled. Let $K$ denote the down sampling rate. After the down sampling, the ECG signal is transmitted. Therefore, we can realize that $T_t=KT_0$.  Essentially, the optimization in this section is performed with respect to $K$ and $W$.

% Assume the practical sampling period is equal to $mT_0$ where $m$ 

% which is related to the oscillator frequency of an analog-digital converter~(ADC). In practice, the sampling at the highest ADC frequency is not necessary. 

%Prior to compressing the ECG data, biomedical sensors sample the ECG signal at the time resolution of $T_0$. The $T_0$ is essentially equal to the minimum time interval over which the ECG signal can be sampled by the sensor. Let $x$ denote the output of the sensor. Therefore, the sensed biomedical signal is 

\subsection{Calculation of Bounds on $T_t$ and $W$}
\label{subsec: bounds}

\subsubsection{Upper bound on $T_t$}
\label{subsubsec: lower b T0}

In general cases, upper bound on sampling period is determined according to Nyquist sampling theorem. For sampling ECG signal, there are some differences.

ECG data is usually taken to assist diagnosis of cardiovascular diseases. In a heart beat period, a ECG signal consists of different waves, such as P wave, QRS wave and T wave. These waves provide assisting information for diagnosing different diseases. For example, ST segment depression or elevation accompanying with T wave inversion is used to diagnose myocardial infarction and cardiogenic shock. QRS voltage, ST-T wave, and R-wave changes are used to diagnose Cardiomyopathy. 

We can easily realize that finer sampling is able to keep more information of ECG waveform. In general cases, durations of different waves are not the same in a ECG waveform. The wave having the smallest duration is most sensitive to sampling period. According to our observations, either Q-R segment or R-S segment has the smallest duration. Time interval between Q and R is denoted by $t_{QR}$. And interval between R and S is $t_{RS}$. To avoid information loss of QRS, we need to guarantee the smaller one between $t_{QR}$ and $t_{RS}$ is larger than the sampling period, $\min\{t_{QR}, t_{RS}\}\leq T_t$. 

%All the information is useful and providing different instructions in heart disease diagnosis.

%Objective of ECG data is to assist diagnosing diseases. One important information ECG is QRS. To avoid losing information of QRS, the three points, Q, R, and S need to be distinguishable. Based on our observations, QRS peaks usually have a relatively more narrow duration than other usual fluctuations. Therefore, duration of QRS peak affects the upper bound on $T_0$.  

As mentioned before, ECG signals significantly changes for different people and different health conditions. Thus, we still use numerical method to investigate the smallest average duration of $t_{QR}$ and $t_{RS}$. In the numerical analysis, we use ECG data published by Michael Oeff~\cite{Goldberger2000} for higher accuracy. The ECG data in~\cite{Goldberger2000} are sampled by 16 bits at the frequency of 10kHz. We estimate the durations of $t_{QR}$ and $t_{RS}$ of ECG data from 549 persons. According to our calculation, average value of $t_{RS}$ is smaller than that of $t_{QR}$. Furthermore, from the 549 recordings, the smallest $t_{RS}$ is equal to 56.9ms. Therefore, we need to guarantee sampling period $T_t$ to provide the time resolution smaller than 56.9ms. Since sampling period of $T_t$ generates resolution of $\frac{T_t}{2}$, the upper bound on $T_t$ is 113.8ms, $T_t^U=0.1138s$.

\subsubsection{Lower Bound on $T_0$}

The finer sampling generates the more accurate ECG signal while more hardware resources are consumed. In the joint optimization on bit rate, smallest sampling period at ADC is considered as the lower bound on $T_0$. In our analysis, $1/360s$ is taken as the low bound on $T_t$, $T_t^L=\frac{1}{360}s$. 

\subsubsection{Upper Bound on $W$}

Larger bit width means finer quantization which provides more accurate description on ECG amplitude. For a given wearable device, bit width $W$ is upper bounded by the implementable largest number of quantization levels. The number of largest bits varies for different wearable devices. We consider 12 as the upper bound on $W$, $W^U=12$. 

\subsubsection{Lower Bound on $W$}

As introduced in Section~\ref{subsubsec: lower b T0}, waves in a beat rate period of ECG signal are used in diagnosing different types of diseases. These waves have different sensitivities to bit number. To determine lower bound on bit width $W$, we need to find out the wave which has the smallest peak average power ratio.  

We still use data from~\cite{Goldberger2000} in the analysis of bit width lower bound. There are four steps in the calculation. 

First, we select the ECG signals in which all waveform features can be observed by a doctor in medicine. The waveform features includes P, Q, R, S and T-waves. 

Next, we measure the waves' summit-to-average distance, which are denoted by $g_{\Xi}$ ($\Xi$ includes an arbitrary member of the alphabet group $\{P, Q, R, S\}$). Essentially, $g_{\Xi}$ is equal to the distance between the locally maximum point of each wave to the base of ECG signals. Companying with each element of $g_{\Xi}$, an envelope amplitude (distance between upper and lower envelope of a ECG recording) is measured. We use $\eta_{\Xi}$ to denote the envelope amplitude companying with $g_{\Xi}$. 

Third, at each ECG recording, we calculate the ratio of $g_{\Xi}$ over its corresponding $\eta_{\Xi}$. The ratio is denoted by $\gamma_k$, $k\in\mathcal{K}$ where $|\mathcal{K}|$ is equal to the number of all calculated ratios. 

Finally, the $\gamma_k$ which has the smallest absolute value is selected to help us determine the lower bound on $W$. Let $W^L$ denote the lower bound. We select $W^L$ such that $\frac{1}{2^{W^U+1}}\leq\min(|\gamma_k|)$. According to our calculation, the lower bound on $W$ is equal to $4$, $W^L=4$.

\subsection{Joint Optimization on Bit Rate}
\label{subsec: elem compr}

%We first consider an element wise compression method which is essentially the quantization of each ECG sample. 

At the wearable device, the quantized ECG data are transmitted to a data server for storage and analysis. We assume the time interval $T_t$ is equal to $KT_0$, $T_t=KT_0$. After receiving the quantized data, the data server reconstructs ECG signal. The reconstructed ECG data is denoted by $\tilde{x}$ which is calculated as follows,
%%%%%%%%%%%%%%%%%%%%%%%%
\begin{equation}
\begin{aligned}
\tilde{x}(mT_0)=\sum_{n=-\infty}^{+\infty}\tilde{x}(nKT_0)
\left[\begin{matrix}u((m-nK)T_0)\\
-u((m-(n+1)K+1)T_0)\end{matrix}\right]. 
\end{aligned}
\label{eq: recon--c}
\end{equation}
%%%%%%%%%%%%%%%%%%%%%%%%%%%%

\begin{comment}
In the compression, there exits the quantization errors which are denoted by $\nu$. Therefore, the connection between arbitrary pair of the points $x(nKT_0)$ and $\tilde{x}(nKT_0)$ is described by
%%%%%%%%%%%%%%%%%%%%%%%%
\begin{equation}
x(nKT_0)=\tilde{x}(nKT_0)+\nu_{nK}.
\label{eq: with quantization error}
\end{equation}
%%%%%%%%%%%%%%%%%%%%%%%%%%%%

The quantization error $\nu_{nK}$ is related to the bit number for quantization ($W$). According to~\cite{Gray1990}, $\nu_{nK}$ uniformly distributes in the range of $[-\frac{1}{2^{W+1}}, \frac{1}{2^{W+1}}]$ when $W$ is large. At the small $W$, the statistical feature of $\nu_{nK}$ is still not known. For an actual ECG compression system, we consider the range of $W$ to be in $[3, 12]$. Through the numerical simulation, we find the quantization noise $\nu_{nK}$ uniformly distributes for $3\leq W\leq 12$.
\end{comment}

With the reconstructed ECG data $\tilde{x}$, we evaluate the reconstruction accuracy in terms of average square error which is denoted by $\varepsilon$. $\varepsilon$ is calculated by
%%%%%%%%%%%%%%%%%%%%%%%%
\begin{equation}
\varepsilon=\lim_{M\rightarrow\infty}\frac{1}{M}\sum_{m=-\frac{M}{2}+1}^{\frac{M}{2}}\left(x(mT_0-\tilde{x}(mT_0))\right)^2.
\label{eq: recon error}
\end{equation}
%%%%%%%%%%%%%%%%%%%%%%%%%%%%

In the ECG data compression and transmission system, the bit rate budget  is $R$ which is essentially a upper bound on the actual bit rate $r$, that is,
%%%%%%%%%%%%%%%%%%%%%%%%
\begin{equation}
r=\frac{W}{T_t}\leq R.
\label{eq: R constraint}
\end{equation}
%%%%%%%%%%%%%%%%%%%%%%%%%%%%

Under the constraint shown in~(\ref{eq: R constraint}), we minimize average square error in reconstructing the ECG signal. The optimization problem is formulated as
%%%%%%%%%%%%%%%%%%%%%%%%
\begin{equation}
\begin{matrix}
\underset{W, K}{\mbox{minimize:}} & \lim_{M\rightarrow\infty}\frac{1}{M}\sum_{m=-\frac{M}{2}+1}^{\frac{M}{2}}\left(c(mT_0-\tilde{c}(mT_0))\right)^2\\
\mbox{subject to:} & \frac{W}{KT_0}\leq R
\label{eq: opt in direct compression}
\end{matrix}
\end{equation}
%%%%%%%%%%%%%%%%%%%%%%%%%%%%

\begin{figure}[h]
\centering
\includegraphics[width=1\linewidth] {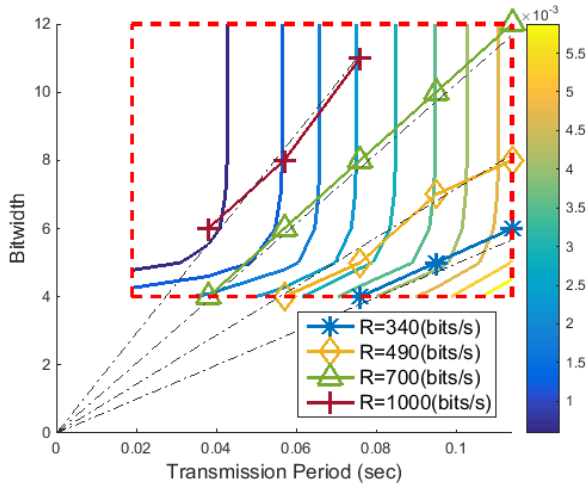}
\setlength{\abovecaptionskip}{3pt plus 3pt minus 2pt}
\caption{ECG reconstruction ASE versus word length in bits and transmission period in second}
\label{fig:direct CP MSE}
%\vspace{-0.15 in}
\end{figure}

In the optimization shown in (\ref{eq: opt in direct compression}), the variables include the average quantization number per sample ($W$) and the transmission period $T_t=KT_0$. Numerical methods are used to solve the optimization problem. Fig.~\ref{fig:direct CP MSE} presents an example of solving optimization problem. 

%%%%%%%%%%%%%%%%%%%%%%%%%%%%%%%%%%%%%%%%%%%%%%%%%%%
%%%%%%%%%%%%%%%%%

In Fig.~\ref{fig:direct CP MSE} the colorful curves are contour of $\varepsilon$. The contours are plotted within a red dash rectangular. The boundaries of the rectangular is formed by the calculated bounds on $W$ and $T_t$. The darker color means the smaller $\varepsilon$. The slope of a dot dash black lines are equal to an ideal bit rate budget $R$. A line with markers indicates the actual bit rate. Since $W$ and $K$ are both in discrete values, the actual bit rate lines with markers can hardly exactly match the ideal lines.

There two steps to determine the minimum MSE under the constraint of $R$. We first draw a bit rate budget line with the slope of $R$. Next, we find the contour curve which is tangent to the bit rate budget line. Then, the contour curve tangent to the budget line informs us the minimum MSE achievable at the bit rate of $R$. 

\section{Experimental Verifications}
\label{sec: experiment}

In this section, experiments are performed to evaluate the effectiveness of the proposed method. We compare our method with existing ones in three aspects, computation complexity, ECG waveform distortion after reconstruction, and the efficiency of compression method in saving bits. Second, we investigate the performance of the conditional quantization at the secondary ECG sensor. Finally, experiments on joint bit rate optimization are performed. 

In the experiments, our objective is to evaluate the effectiveness of proposed ECG compression method. In our knowledge, the compression is not as sensitive to the change of waveform shapes as the algorithms for R-wave detection or other cardiovascular disease diagnosis. Therefore, the experiments are performed on the data from only two databases, MIT-BIH Arrhythmia database~\cite{Goldberger2000} and European ST-T database~\cite{Goldberger2000a}.

\subsection{Complexity and Reconstruction Accuracy Comparison}

In this subsection, we investigate the performance of the proposed differential ECG compression method. To evaluate the performance, we compare our method with the ones in literature. The ECG compression schemes based on DCT and wavelet are considered since they are widely adopted in ECG compression. Different from pure wavelet algorithm, wavelet compression by the set partitioning in hierarchical trees algorithms is implemented. Since compressive sensing is widely discussed and applied, ECG signal is compressed via a compressive sensing algorithm. We also evaluate performance of two differential ECG compression methods, least mean square~(LMS) based compression algorithm and DPCM based one.

%%%%%%%%%  Table 1  %%%%%%%%%%%%%
\begin{table}[t]
\centering
\caption{Comparison in computation complexity}
\begin{tabular}{cccc}
\hline\hline
& \specialcell{Mul. \\(/sample)} & \specialcell{Addl. \\(/sample)}  & Memory U. \\
\hline
 LMS & 4 & 5  & 8\\
LMS (no. coef.) & 4 & 5  & 8\\
DCT & 39 & 1 & 1601\\
Wavelet+SPIHT & 68 & 143 & 1216 \\
Delta modulator & 1 &  2 & 1 \\
Compressive Sensing & 212 & 183 & 2022\\
New method & 0  & 3 & 2\\
\hline
\end{tabular}
\label{tb: computation complexity}
\end{table}
%%%%%%%%%%%%%%%%%%%%%%%

Table~\ref{tb: computation complexity} presents computation complexities of different ECG compression methods. Average numbers of multiplications, additions per data sample and required memory units are taken as the metrics. From Table~\ref{tb: computation complexity}, DCT and wavelet based compressors need a large number of multiplications and memory units. The large number of multiplications are induced by the multiplication between ECG signal vector and groups of basis vectors. The volume of required multiplications is related to the length of a ECG segment. Besides the extensive demand on multipliers, a large number of memory units are also needed. LMS based compressor has lower computation complexity than the previous two compressors. From Table~\ref{tb: computation complexity}, we can easily find that both DPCM based compressor and our method can be implemented in low complexity. Different from Delta modulator based compressor, our method does not need multiplication operations.

 %%%%%%%%%%%%%%%%%%

%\subsection{Simulation in Bit Rate Optimization}

\begin{figure}[h]
\centering
\includegraphics[width=1\linewidth] {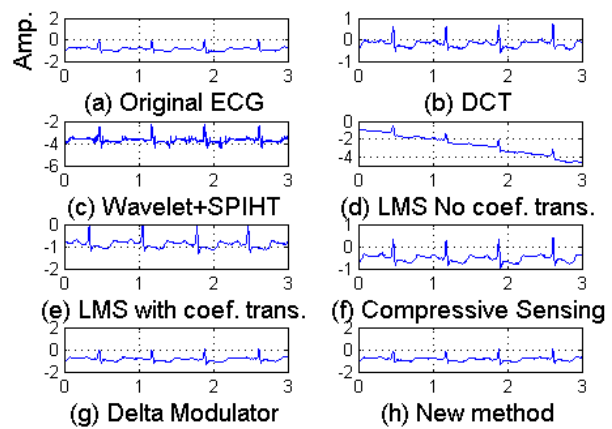}
\setlength{\abovecaptionskip}{3pt plus 3pt minus 2pt}
\caption{Comparison of the reconstructions of No. 112 record in MIT-BIH arrhythmia database}
\label{fig:ECG 112}
%\vspace{-0.15 in}
\end{figure}

\begin{figure}[h]
\centering
\includegraphics[width=1\linewidth] {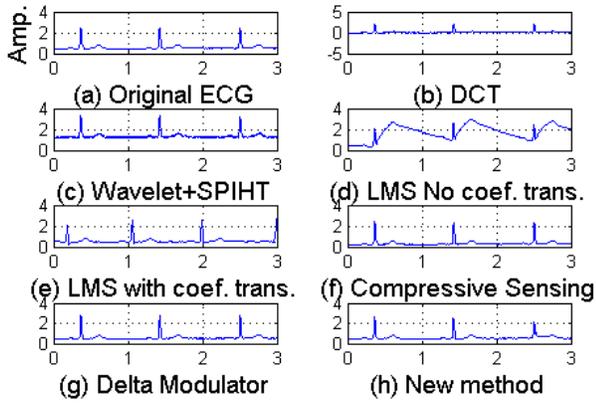}
\setlength{\abovecaptionskip}{3pt plus 3pt minus 2pt}
\caption{Comparison of the reconstructions of No. 103 record in European ST-T database}
\label{fig:ECG 103 eu}
%\vspace{-0.15 in}
\end{figure}

To present an intuitive impression on the performance of compression algorithms, we present the reconstructed ECG waveforms by all the mentioned algorithms. Due the page limits, the graphic performance comparison is performed on two ECG records, No. 112 record in MIT-BIH database and No. 103 record in European ST-T database. The reconstruction accuracy comparison for the two records are presented in Fig.~\ref{fig:ECG 112} and Fig.~\ref{fig:ECG 103 eu} respectively. The computation is the compressions are performed in 8-bits numbers.

%%%%%%%%%%%%%%%%%%%%%%%%%%%%
%%%%%%%%%%%%%%%%%

%%%%%%%%%%%%%%%%%%%%%%%%%%%%%
%%%%%%%%%%%%%%%%%

From Fig.~\ref{fig:ECG 112}, the reconstruction ECG via the DCT based method retains the key features, such as P, Q, R, S, and T waves. Wavelet based compression incurs some noise-like distortion. In Fig.~\ref{fig:ECG 112} (d), coefficients of adaptive filter are not updated and we can observe ECG waveform distortion. In Fig.~\ref{fig:ECG 112} (e), the coefficients are adaptively updated which generate satisfying reconstruction accuracy. However, the bits used for updating coefficients are in a large number. The accuracy of the reconstructed ECG signal from the compressive sensing method is high while the computation burden is heavy. The key features of ECG signal can also be observed in DPCM based compressor (Fig.~\ref{fig:ECG 112} (g)). However, we can find unexpected fluctuations between R and S. DPCM quantizes the error occurring estimating current ECG value. When ECG waveform changes fast, such as in the segment between R and S, DPCM is not able to keep tracking of the fast change. Thus, the unexpected fluctuations occur. Our method quantizes the ECG amplitude change directly. Thus, our method is more robust to the fast change. The similar phenomena can also be observed in Fig.~\ref{fig:ECG 103 eu}.

%, except the one in Fig.~\ref{fig:ECG 112} (c). From Fig.~\ref{fig:ECG 112} (c), if coefficients in LMS are not updated, reconstructed ECG waveform is distorted and the distortion keep being amplified with time. This is caused by accumulation of quantization error in each step of LMS filtering. 

%From Fig.~\ref{fig:ECG 106}, as for DCT based compression method, reconstructed ECG signal matches the one before compression. We can clearly observe the key features from the reconstructed ECG signal, such as P, Q, R, S, and T waves. Wavelet based compression results have obvious waveform distortion. In Fig.~\ref{fig:ECG 106} (c), coefficients of adaptive filter are not updated and we can observe ECG waveform distortion. In Fig.~\ref{fig:ECG 106} (d), the coefficients are adaptively updated which generate better reconstruction accuracy. The key features of ECG signal can also be observed in DPCM based compressor (Fig.~\ref{fig:ECG 106} (e)). However, we can find unexpected fluctuations between R and S. The fluctuation caused waveform distortion is not observed in our method. DPCM quantizes the error occurring estimating current ECG value. When ECG waveform changes fast, such as in the segment between R and S, DPCM is not able to keep tracking of the fast changing. Thus, the unexpected fluctuations occur. Our method quantizes the ECG amplitude change directly. Thus, our method is more robust to the fast changing. 

%%%%%%%%%%%%%%%%%%%%%%%%%%%%%%%%%%%%%%%%%%%%%%%%%%%
\begin{figure}[h]
\centering
\includegraphics[width=1\linewidth] {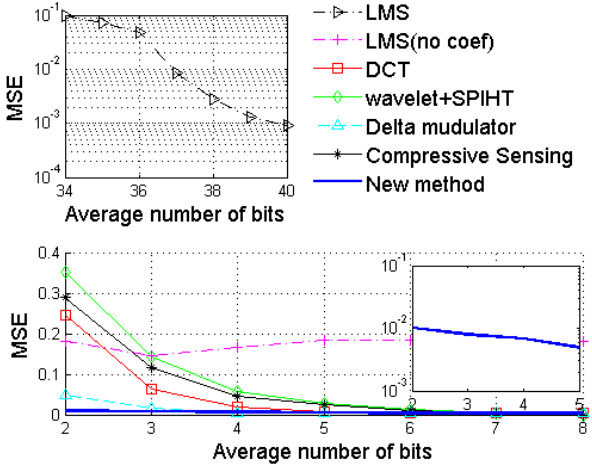}
\setlength{\abovecaptionskip}{3pt plus 3pt minus 2pt}
\caption{MSE versus average number of bits per sample for records in MIT-BIH arrhythmia database}
\label{fig:R-D fixed T0}
%\vspace{-0.15 in}
\end{figure}
%%%%%%%%%%%%%%%%%

%%%%%%%%%%%%%%%%%%%%%%%%%%%%%%%%%%%%%%%%%%%%%%%%%%%
\begin{figure}[h]
\centering
\includegraphics[width=1\linewidth] {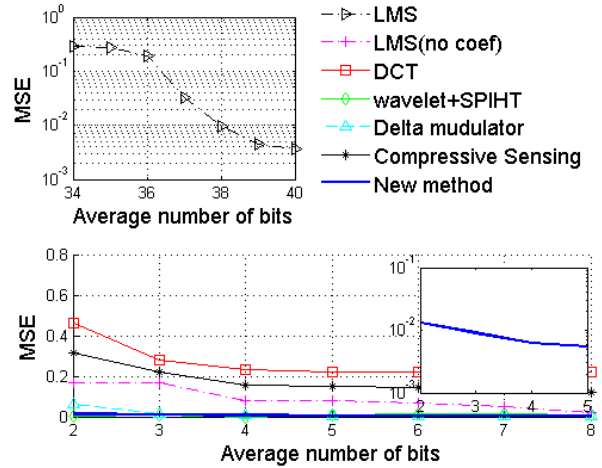}
\setlength{\abovecaptionskip}{3pt plus 3pt minus 2pt}
\caption{MSE versus average number of bits per sample for records in European ST-T database}
\label{fig:R-D fixed T0 Europe}
%\vspace{-0.15 in}
\end{figure}
%%%%%%%%%%%%%%%%%

Fig.~\ref{fig:ECG 112} and Fig.~\ref{fig:ECG 103 eu} illustrate the reconstructed ECG waveform at a fixed bit width of 8. Furthermore, we present the normalized MSE of the reconstructed ECG at different bit widths in Fig.~\ref{fig:R-D fixed T0} and Fig.~\ref{fig:R-D fixed T0 Europe} which curves are calculated from the records in MIT-BIH Arrhythmia database and European ST-T database respectively. From Fig.~\ref{fig:R-D fixed T0}, wavelet based compression induces the worst reconstruction accuracy. When 4 coefficients of LMS filters are not transmitted from a compressor, MSE does not decrease with bit number increasing. When the coefficients are transmitted, MSE decreases significantly with bit width increasing. However, the coefficient updating requires more bits. Our new method achieve the smallest MSE at a given low bit rate. The advantage of the new method over the existing ones can also be observed in Fig.~\ref{fig:R-D fixed T0 Europe}.

\subsection{Simulation in Double Sensors Based ECG Compression}

In this subsection, we investigate the performance of double sensors based ECG compression method. As discussed in previous sections, the distribution of ECG signal from primary sensor is priorly known by the secondary sensor. Thus, conditional quantization can be performed at the secondary sensor. The quantization results are transmitted to the primary sensor via a perfect channel. The primary sensor differentially quantize ECG signal acquired by itself. The results of conditional quantization and differential quantization are transmitted to a remote data center. At the data center, the ECG signal acquired by the primary sensor is first reconstructed. Then, reconstruction of the ECG signal from the secondary sensor is performed.

%%%%%%%%%%%%%%%%%%%%%%%%%%%%%%%%%%%%%%%%%%%%%%%%%%%
\begin{figure}[h]
\centering
\includegraphics[width=1\linewidth] {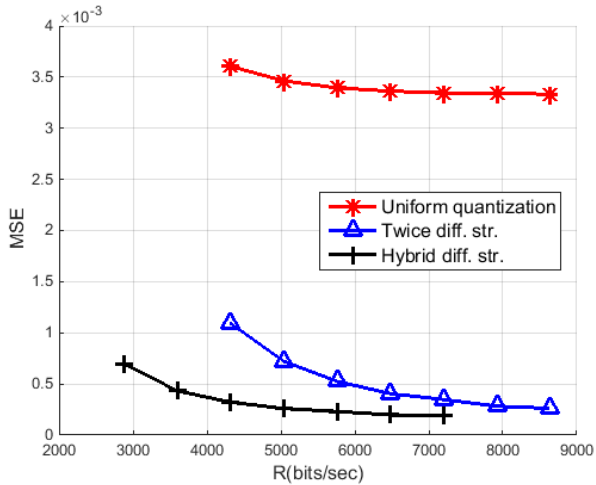}
\setlength{\abovecaptionskip}{3pt plus 3pt minus 2pt}
\caption{Bit rate in ECG compression on double sensors}
\label{fig: double compress}
%\vspace{-0.15 in}
\end{figure}
%%%%%%%%%%%%%%%%%

The combination of differential quantization at primary sensor and the conditional quantization at the secondary sensor is called as hybrid quantization structure. For comparison, we also consider other two quantization structures in the double sensors based compression. First, differential quantization is applied in both primary and secondary sensors. Second, differential quantization is taken at the primary sensor and uniform quantization used at the secondary one. The average MSE of the reconstructed ECG at both the primary and secondary sensors is taken as the accuracy metric. 

The results of the two sensors based ECG compression are presented in Fig.~\ref{fig: double compress}. From the results, the proposed hybrid quantization method outperforms other two, the conventional quantization on original ECG signal (labeled as 'uniform quantization') and the two-independent-differential ECG quantization (labeled as 'twice diff. str'). The advantage of the proposed two sensors based compression scheme is caused by the reduction of redundant information between ECG signal from the two sensors.

\subsection{Simulation in Joint Bit Rate Optimization}

Until now, the simulations are performed in the precondition that the sampling period is fixed. In this subsection, we investigate the performance of joint bit rate optimization over quantization bit number and transmission period. As a comparison, quantization results are transmitted at a period of $1/360s$ which is the same with the sampling period for the records in MIT-BIH database. Via interpolation, the equivalent sampling at $360Hz$ is also performed on the data records for European ST-T database.

%%%%%%%%%%%%%%%%%%%%%%%%%%%%%%%%%%%%%%%%%%%%%%%%%%%
\begin{figure}[h]
\centering
\includegraphics[width=1\linewidth] {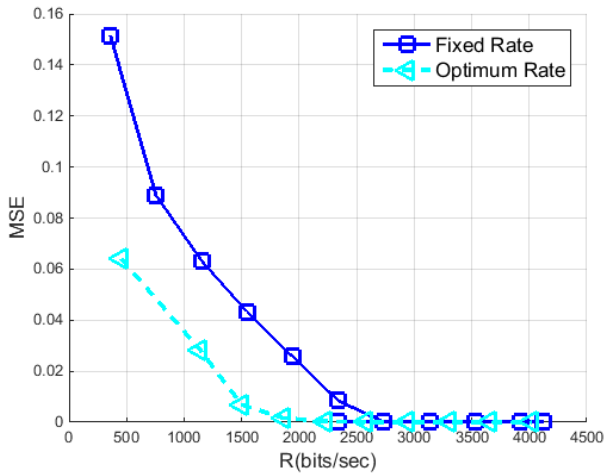}
\setlength{\abovecaptionskip}{3pt plus 3pt minus 2pt}
\caption{Joint bit rate optimization}
\label{fig: joint rate opt}
%\vspace{-0.15 in}
\end{figure}
%%%%%%%%%%%%%%%%%

%In Fig.~\ref{fig:R-D fixed T0}, \enquote*{MSE-bit rate} curves are generated at fixed sampling period. Indeed, the joint optimization over bit number and transmission period is able to reduce bit rate consumption. 

Fig.~\ref{fig: joint rate opt} plots \enquote*{MSE-bit rate} curves with and without the joint optimization. The two dimensional optimization is constrained by the bounds both in sampling period and quantization bits per sample. These bounds, which are calculated in Section~\ref{subsec: bounds}, enable us guarantee key features of ECG signal can be retained in the compression. From Fig.~\ref{fig: joint rate opt}, advantage of the joint optimization on saving bit rate can be clearly observed. 

\section{Conclusion}
\label{sec:conclusion}

We investigate the compression of ECG signal which is important for saving hardware and power consumption in health telemonitoring systems. Different from the ECG compression work in literature, compression scheme based on multiple ECG sensors is considered. Without loss of generality, we consider an example in which there are two ECG sensors, a primary and a secondary ECG sensor. At the primary one, we use a novel differential structure to compress ECG signal which effectively reduces the redundant information between adjacent ECG samples. At the secondary ECG sensor, conditional quantizer is proposed to compress ECG signal which utilizes the inherent connection between the shapes of ECG signals from the two sensors. Experiments verify the advantage of our proposed compression scheme both in complexity and reconstruction accuracy.

\linespread{1.27}
\bibliography{chann_emu_InfoCom2013}
\bibliographystyle{IEEEtran}

\end{document}